\documentclass[pra,aps,superscriptaddress,twocolumn,showpacs]{revtex4}

\usepackage{graphicx}
\usepackage[normal]{subfigure}
\usepackage{caption}
\usepackage{latexsym}
\usepackage{amsmath}
\usepackage{amssymb}
\usepackage{amsfonts}
\usepackage{color}

\begin{document}

\newcommand{\ket}[1]{\vert #1 \rangle}
\newcommand{\bra} [1] {\langle #1 \vert}
\newcommand{\braket}[2]{\langle #1 | #2 \rangle}
\newcommand{\proj}[1]{\ket{#1}\bra{#1}}
\newcommand{\mean}[1]{\langle #1 \rangle}
\newcommand{\opnorm}[1]{|\!|\!|#1|\!|\!|_2}
\newtheorem{theo}{Theorem}
\newtheorem{lem}{Lemma}


\title{Interconversion of pure Gaussian states using non-Gaussian operations}

\author{Michael G. Jabbour}
\author{Ra\'{u}l Garc\'{\i}a-Patr\'{o}n}
\author{Nicolas J. Cerf}
\affiliation{Quantum Information and Communication, Ecole Polytechnique de Bruxelles, CP 165, Universit\'e libre de Bruxelles,
1050 Bruxelles, Belgium}


\begin{abstract}
We analyze the conditions under which local operations and classical communication enable entanglement transformations within the set of bipartite pure Gaussian states. A set of necessary and sufficient conditions had been found in [Quant. Inf. Comp. \textbf{3,} 211 (2003)] for the interconversion between such states that is restricted to Gaussian local operations and classical communication. Here, we exploit majorization theory  in order to derive more general (sufficient) conditions for the interconversion between bipartite pure Gaussian states
that goes beyond Gaussian local operations. While our technique is applicable to an arbitrary number of modes for each party, it allows us to exhibit surprisingly simple examples of $2\times 2$ Gaussian states that necessarily require non-Gaussian local operations to be transformed into each other.
\end{abstract}


\maketitle


\section{Introduction}

Quantum entanglement plays a major role in quantum computation and information theory, where it is a key resource enabling a vast variety of tasks, such as teleportation-based universal quantum computing \cite{GottesmanChuang} or blind quantum computing \cite{Barz}, and where it is also a central component of the security analysis of quantum key distribution \cite{Scarani}. More fundamentally, understanding quantum entanglement is at the heart of theoretical physics, with issues ranging from quantum non-locality and Bell inequalities \cite{Brunner} to novel perspectives on black holes theory \cite{Adami,Maldacena}.

A fundamental problem in the theory of quantum entanglement consists of classifying entangled states into different equivalence classes and studying the possibility (or impossibility) of transforming entangled states between each other \cite{Horodecki}. In the usual scenario, one considers the interconversion between different sets of states that rely on local operations (effected separately by each of the parties sharing the state) supplemented with classical communication (among the different parties sharing the state). These transformations, commonly denoted as LOCC, cannot increase the entanglement between the parties whatever the entanglement monotone used to measure this entanglement. It is, however, crucial to go beyond that simple fact and be able to determine whether a given entangled state can be reached by applying a LOCC transformation onto another entangled state.
A very successful approach to address this question in the case of bipartite pure entangled states in finite dimension has been developed based on the mathematical theory of majorization:
the possibility to transform a pure bipartite state into another by a deterministic LOCC protocol is connected to a majorization relation between their corresponding vectors of Schmidt coefficients \cite{Nielsen,NielsenVidal}.

In this article, we envisage the interconversion between states of the electromagnetic field, and move therefore to an infinite-dimensional Fock space. We focus in particular on the set of Gaussian states, which are of great significance in quantum optics and continuous-variable quantum information theory \cite{BraunsteinvanLoock,CerfLeuchsPolzik}. These states are easy to produce and manipulate experimentally, and at the same time can be described mathematically by using the first two statistical moments of the quadrature operators in phase space. Consequently, the set of Gaussian states and accompanying Gaussian transformations is particularly relevant for our analysis of entanglement transformations as they
well describe a great amount of quantum optical experiments and can be efficiently modelled within the so-called symplectic formalism \cite{Weedbrook2011}.

The interconversion between Gaussian states has been addressed in a few earlier works \cite{EisertPlenio,Eisert-nogo,Giedke-nogo,Fiurasek,Giedke,GiedkeKraus}, but many problems remain unsolved. In particular, most works have focused on the entanglement transformations of Gaussian states using Gaussian processes only. Notably, the work  by Giedke \textit{et al.} \cite{Giedke}  provides a necessary and sufficient condition for the interconversion between pure Gaussian states when restricting to Gaussian local operations with classical communication (denoted as GLOCC), which can be realized in practice by using standard optical components, such as beam splitters, squeezers, phase-shifters and homodyne detectors. The proof starts by exploiting the fact that any $N \times N$ pure Gaussian state, i.e., any bipartite Gaussian state with $N$ modes on each side, can be transformed by a Gaussian local unitary into a tensor product of $N$ two-mode squeezed vacuum states \cite{Weedbrook2011}, which is completely characterized by the vector 
of squeezing parameters $\mathbf{r^{\downarrow}}$ conventionally sorted in decreasing order. Then, the problem simplifies to the interconversion between tensor products of two-mode squeezed vacuum states. The authors proved that a pure Gaussian state $\ket{\psi}$ can be transformed into $\ket{\psi'}$ using a GLOCC if and only if $r_i \geq r'_i$, $\forall i$, or in a more compact notation
\begin{equation}
\ket{\psi} \xrightarrow{\text{GLOCC}} \ket{\psi'} \text{  ~~iff~~ } \boldsymbol{r^{\downarrow}} \geq \boldsymbol{r'^{\downarrow}},
\label{eqGiedke}
\end{equation}
where $\ket{\psi}$ and $\ket{\psi'}$ are respectively characterized by their decreasing-ordered squeezing vectors $\boldsymbol{r^{\downarrow}}$ and $\boldsymbol{r'^{\downarrow}}$.

The question that we investigate in the present work is whether it is possible to achieve, using a non-Gaussian LOCC, transformations between
Gaussian states that are otherwise inaccessible by a GLOCC, i.e., do not satisfy condition (\ref{eqGiedke}).
This possibility was briefly mentioned in Ref.~\cite{Giedke} with the example of a couple of Gaussian states that could be connected via a LOCC but could not via a Gaussian process alone. Here, we extend on this idea and develop a systematic approach to explore the possible interconversion between Gaussian states that are not accessible by GLOCC. We broaden the analysis by providing a sufficient condition for the existence of a LOCC transformation between pure bipartite $n\times n$ Gaussian states that generalizes condition (\ref{eqGiedke}), at the price of loosing its necessary character.

To achieve this goal, we use the theory of majorization, which
provides an ideal tool to investigate the conditions for interconverting pure bipartite states using LOCC transformations \cite{Nielsen,NielsenVidal}.
Specifically, a finite-dimensional bipartite pure state $\ket{\psi}$ can be transformed via a deterministic LOCC into $\ket{\psi'}$ if and only if the vectors of eigenvalues $\boldsymbol{\lambda}$ and $\boldsymbol{\lambda'}$ of 
their respective reduced states $\rho$ and $\rho'$ satisfy a majorization relation, that is 
\begin{equation}
\ket{\psi} \xrightarrow{\text{LOCC}}  \ket{\psi'}  \text{  ~~iff~~ } \boldsymbol{\lambda'}\succ\boldsymbol{\lambda},
\end{equation}
Equivalently, the interconversion $\ket{\psi} \to \ket{\psi'}$ is possible if and only if
there is a bistochastic matrix $\mathbf{D}$ that maps $\boldsymbol{\lambda'}$ onto $\boldsymbol{\lambda}$, i.e., $\boldsymbol{\lambda}=\mathbf{D}\boldsymbol{\lambda'}$,
as reviewed in the Appendix. We will actually use this second condition in the present paper, as it is better appropriate to achieve our goal (finding a matrix $\mathbf{D}$ gives a sufficient condition for the existence of a LOCC transformation).

The application of majorization theory in an infinite-dimensional Fock space in order to explore the interconversion between relevant states in quantum optics is a fertile ground of investigation, with only a few known results as of today.  In Ref.~\cite{Raul}, it was shown that the output states of an optical quantum-limited amplifier that is applied to Fock states satisfy a ladder of majorization relations, while a similar behaviour was proven to hold in Ref.~\cite{Christos} for a pure lossy line (a beam splitter with vacuum on the other input port). Besides these majorization relations intrinsic to these generic two-mode Gaussian operations (two-mode squeezer and beam splitter), it was recently shown that the output state resulting from the vacuum state processed by any phase-insensitive Gaussian bosonic channel majorizes the output state corresponding to another input state \cite{Mari}, extending on the proof of the Gaussian minimum entropy conjecture \cite{Giovannetti}. The results presented here follow this line and illustrate again the power of majorization theory in quantum optics and continuous-variable quantum information theory.


The remaining of the manuscript is organized as follows. In Section II, we outline the main results, illustrated in the simplest case of pure Gaussian states of $2 \times 2$ modes.
In Section III, we give the proof of our main theorem, as well as its generalization to an arbitrary number of modes.
Finally, in Section IV, we conclude and bring forward some open problems. 
In the Appendix, we summarize the basics of majorization theory and its connection to entanglement theory.


\section{Interconversion of Gaussian states of $2\times 2$ modes}

We first illustrate our results for the simplest interesting case, namely pure Gaussian states of $2 \times 2$ modes. Indeed, the case of pure Gaussian states of $1 \times 1$ modes trivially reduces to the interconversion between two-mode squeezed vacuum states,
\begin{equation}
|\Phi_r\rangle = (1-\gamma^2)^{1/2} \sum_{n=0}^\infty \gamma^n |n\rangle |n\rangle
\end{equation}
with $\gamma=\tanh(r)$ and $r$ being the squeezing parameter. An obvious consequence of condition (\ref{eqGiedke}) is that $|\Phi_r\rangle$ can be transformed into $|\Phi_{r'}\rangle$ with a GLOCC iff $r\ge r'$. If $r<r'$ the transformation is possible in the opposite direction with a GLOCC, so non-Gaussian transformations are useless both ways. The situation becomes more interesting as soon as 2 modes are considered on each side.
Since any pure Gaussian state of $2 \times 2$ modes can be mapped via Gaussian local unitaries onto a pair of two-mode squeezed vacuum states \cite{Weedbrook2011}, we can assume without loss of generality that Alice and Bob's state $\ket{\psi}$ is already in its normal form $|\Phi_{r_1}\rangle |\Phi_{r_2}\rangle$, characterized by its ordered squeezing vector $(r_1,r_2)$, with $r_1\ge r_2$. We then study the conditions under which it can be transformed under a LOCC into another state $\ket{\psi'}=|\Phi_{r_1'}\rangle |\Phi_{r_2'}\rangle$ with squeezing vector $(r'_1,r'_2)$ and $r_1'\ge r_2'$.

\begin{figure}[b]
\centering
\subfigure[]{\label{a}\includegraphics[width=14mm]{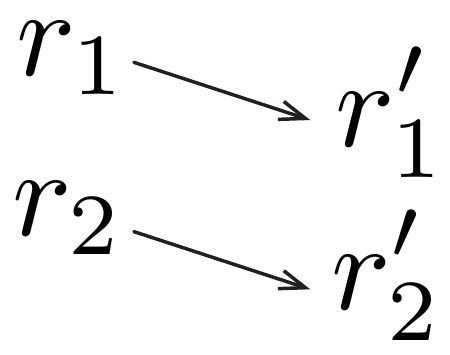}}
\hspace{1.5em}
\subfigure[]{\label{b}\includegraphics[width=14mm]{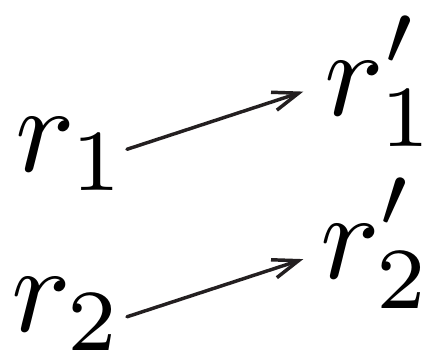}}
\hspace{1.5em}
\subfigure[]{\label{c}\includegraphics[width=14mm]{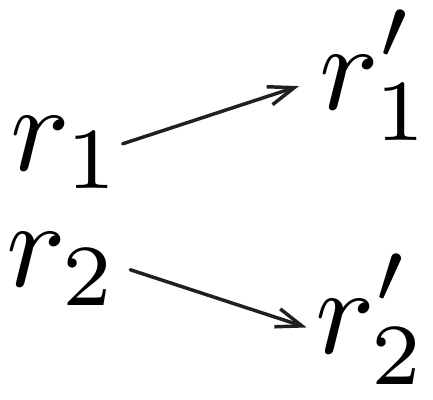}}
\hspace{1.5em}
\subfigure[]{\label{d}\includegraphics[width=14mm]{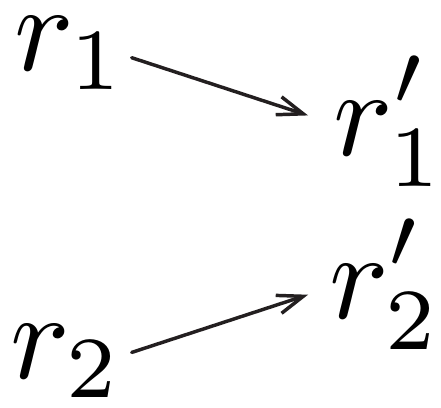}}
\caption{Four possible evolutions of the squeezing parameters $r_1$ and $r_2$ of the two two-mode squeezed vacuum states $|\Phi_{r_1}\rangle |\Phi_{r_2}\rangle$. Note that since the squeezing vectors are sorted in decreasing order ($r_1\ge r_2$, $r_1'\ge r_2'$), the two arrows never cross each other.}
\label{fig1}
\end{figure}

As depicted in Figure \ref{fig1}, there are four different possibilities for the evolution of the squeezing vector's components when transforming $\ket{\psi}$ into $\ket{\psi'}$.
In case \ref{a}, both squeezing parameters $r_1$ and $r_2$ decrease, a transformation that is always achievable using a GLOCC according to condition \eqref{eqGiedke}.
Case \ref{b} corresponds to the situation where both squeezing parameters $r_1$ and $r_2$ increase.
It is easy to infer that there cannot exist a LOCC (neither Gaussian nor non-Gaussian) that permits such a transformation. Indeed,
case \ref{b} can be viewed as the reverse process of case \ref{a} and condition (\ref{eqGiedke}) permits a GLOCC in the reverse direction. Hence, transformation \ref{b} is forbidden as majorization is a one-way property. 
In the latter two cases, the squeezing parameters follow different evolutions, i.e., the two components of the vector $(r_1'-r_1,r_2'-r_2)$ have opposite signs. 
In case \ref{c}, $r_1$ increases and $r_2$ decreases, while case \ref{d} is the converse. Since condition \eqref{eqGiedke} is necessary, it is never possible to transform $\ket{\psi}$ into $\ket{\psi'}$ using a GLOCC in neither of these cases. However, as it turns out, it is nevertheless possible to find a non-Gaussian LOCC that successfully achieves such a transformation when the condition of the following theorem holds:

\begin{theo}
Let $\ket{\psi}$ and $\ket{\psi'}$ be two $2 \times 2$ pure Gaussian states respectively characterized by their decreasingly ordered squeezing vectors $(r_1,r_2)$ and $(r_1',r_2')$ such that the two components of vector $(r_1'-r_1,r_2'-r_2)$ have opposite signs. Then, $\ket{\psi}$ can be transformed into $\ket{\psi'}$ using a non-Gaussian LOCC if
\begin{equation}
\frac{\sinh(r_1+r_2) \pm  \sinh(r_1-r_2)}{\sinh(r'_1+r'_2) \pm \sinh(r'_1-r'_2)} \geq 1,
\label{eqMain}
\end{equation}
where $\pm$ follows the sign of $r_1'-r_1$. [The plus sign corresponds to case \ref{c}, while the minus sign corresponds to case \ref{d}.]
\label{MainResult}
\end{theo}

\begin{figure*}
\includegraphics[width=14cm]{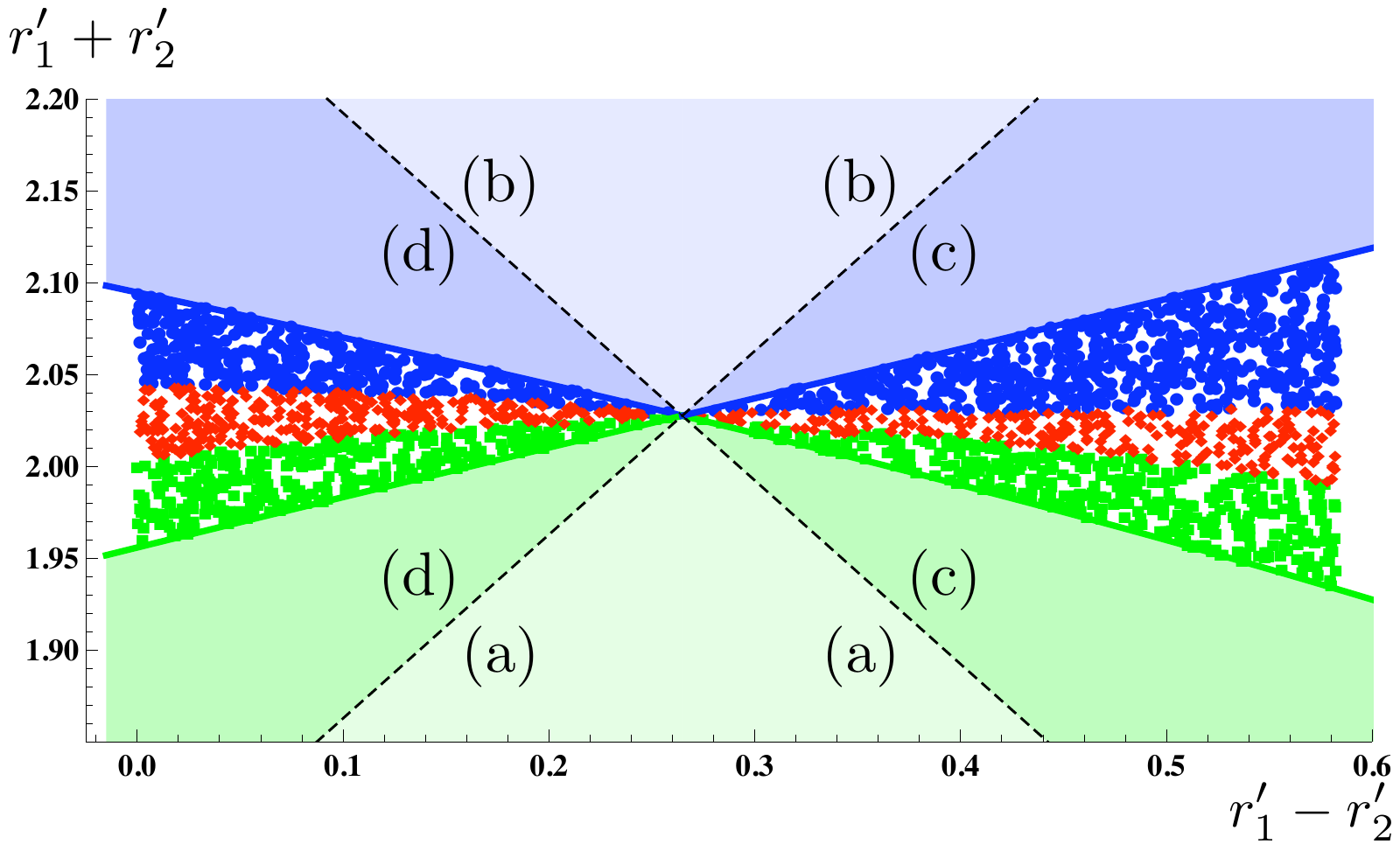}
\caption{Transformation from state $\ket{\psi}=|\Phi_{r_1}\rangle |\Phi_{r_2}\rangle$ to state $\ket{\psi'}=|\Phi_{r_1'}\rangle |\Phi_{r_2'}\rangle$ for fixed values $r_1=1.15$ (9.96~dB), $r_2= 0.88$ (7.66~dB), and variable $(r_1',r_2')$. The two diagonal dashed lines divide the plane in four quadrants corresponding to the four situations depicted in Fig.~\ref{fig1}. The final states that are reachable with a LOCC transformation are marked in green (green areas and green points), while the final states  that are not reachable because a LOCC transformation exists in the opposite direction are marked in blue (blue areas and blue points). The final states that are incomparable with the initial states (no LOCC exists both ways) are marked as red points. The lower quadrant (light green) corresponds to case \ref{a}, while the upper quadrant (light blue) corresponds to case \ref{b}. Condition (\ref{eqGiedke}) implies that a GLOCC exists in the lower quadrant, while a LOCC is ruled out in the upper quadrant since a GLOCC exists in the reverse direction. The right quadrant [case \ref{c}] and left quadrant [case \ref{d}] correspond to regions where condition (\ref{eqGiedke}) is not satisfied both ways, so no GLOCC exists both ways. However, non-Gaussian LOCC are possible and detected by our condition (\ref{eqMain}) in the dark green region (or non-detected by our condition and represented by green points). Symmetrically, non-Gaussian LOCC are possible in the reverse direction and detected by our condition (\ref{eqMain}) in the dark blue region (or non-detected by our condition and represented by blue points); this corresponds to cases where no LOCC exists that converts $\ket{\psi}$ into $\ket{\psi'}$. Our criterion, condition (\ref{eqMain}), is indicated by a solid green line (direct direction $\ket{\psi}\to\ket{\psi'}$) and blue line (reverse direction $\ket{\psi'}\to\ket{\psi}$).}
\label{fig2}
\end{figure*}


Thus, Theorem \ref{MainResult}  provides us with a sufficient (but not necessary) condition for the existence of a LOCC that transforms $\ket{\psi}=|\Phi_{r_1}\rangle |\Phi_{r_2}\rangle$  into  $\ket{\psi'}=|\Phi_{r_1'}\rangle |\Phi_{r_2'}\rangle$. The usefulness of this result is illustrated in Fig.~\ref{fig2}, where we analyse the possibility of the conversion from the initial state $\ket{\psi}$ to the final state $\ket{\psi'}$ for fixed values of the squeezing parameters of the initial state,  $r_1=1.15$ (9.96~dB) and $r_2= 0.88$ (7.66~dB). These values correspond simply to a mean photon number of 2 and 1, respectively, when tracing $|\Phi_{r_1}\rangle$ and $|\Phi_{r_2}\rangle$ over Bob's mode.
Each possible final state with squeezing parameters $r_1'$ and $r_2'$ is associated with a point of coordinates $(r_1'-r_2',r_1'+r_2')$ in Fig.~\ref{fig2}. Thus, we can divide the plane in four quadrants corresponding to the four cases of Fig.~\ref{fig1}, where the increasing-diagonal dashed line coincides with $r_2'=r_2$ while the decreasing-diagonal dashed line corresponds to $r_1'=r_1$. The lower quadrant (light green) corresponds to case \ref{a}, i.e., transformations that are achievable with a GLOCC according to condition (\ref{eqGiedke}). The upper quadrant (light blue) corresponds to case \ref{b}, i.e., transformations that cannot be achieved with a LOCC since condition (\ref{eqGiedke}) permits a GLOCC in the reverse direction. The right quadrant corresponds to case \ref{c}, while the left quadrant corresponds to case \ref{d}. In both left and right quadrants, the transformations cannot be achieved with a GLOCC as they do not satisfy condition (\ref{eqGiedke}), but might be achievable with a non-Gaussian LOCC. Indeed, condition (\ref{eqMain}) is satisfied in a whole region (dark green) above the lower quadrant, implying that a LOCC exists, which must necessarily be non-Gaussian. Symmetrically, condition (\ref{eqMain}) is satisfied in the reverse direction $\ket{\psi'} \longrightarrow \ket{\psi}$ in a whole region (dark blue) below the upper quadrant, which means that a LOCC (be it non-Gaussian) exists in the reverse direction, hence the transformation $\ket{\psi} \longrightarrow \ket{\psi'}$ is impossible with a LOCC. Thus, the sufficient condition (\ref{eqMain}) allows us to significantly enlarge the regions where a LOCC is proven to exist (green area) or not to exist (blue area). This is the main outcome of Theorem \ref{MainResult}.

Now, the remaining zones in the left and right quadrants can be explored numerically. For each point, we check whether a majorization relation exists between the eigenvalues of the reduced states corresponding to $\ket{\psi}$ and $\ket{\psi'}$ in either direction, or whether the states are incomparable. We find states (marked as green points) where majorization holds, so that a LOCC exists although it is not detected by condition (\ref{eqMain}), illustrating the fact that this condition is not necessary. Similarly, we find states (marked as blue points) such that majorization holds in the opposite direction (undetected by our condition), implying that a LOCC exists in that direction, hence the transformation $\ket{\psi} \longrightarrow \ket{\psi'}$ is impossible with a LOCC. Finally, we find states that are incomparable (marked as red points), in which case no deterministic LOCC exists both ways.

\section{Proof and extension to Gaussian states of $N\times N$ modes}

In general terms, we are interested in deterministic LOCC transformations from the pure Gaussian state $\ket{\psi}_{AB}$, shared between Alice having $N$ modes and Bob having $N$ modes, towards the pure Gaussian state $\ket{\psi'}_{AB}$. Using the normal form reduction \cite{Weedbrook2011}, we can assume without loss of generality that both $\ket{\psi}_{AB}$ and $\ket{\psi'}_{AB}$ are a tensor product of two-mode squeezed vacuum states characterized by their respective squeezing vectors $\boldsymbol{r}$ and $\boldsymbol{r'}$. So, we seek a sufficient condition on the existence of transformation
\begin{equation}
\ket{\psi}_{AB} = |\Phi_{r_1}\rangle \cdots |\Phi_{r_N}\rangle \rightarrow \ket{\psi'}_{AB} = |\Phi_{r_1'}\rangle \cdots |\Phi_{r_N'}\rangle
\end{equation}
under a deterministic LOCC.
As shown for example in \cite{Raul}, majorization theory and its connection to entanglement transformations can be adapted to infinite dimensions and the usual states of quantum optics, the only subtlety being that the matrix $\mathbf{D}$ becomes an infinite column-stochastic matrix instead of a double-stochastic matrix (this means that all columns must still sum to one, while the sum of elements in each row must only be $\le 1$). Majorization theory implies that a pure state $\ket{\psi}_{AB}$ can be transformed into
$\ket{\psi'}_{AB}$ using a deterministic LOCC if and only if Alice's reduced states $\rho={\rm Tr}_B[\proj{\psi}_{AB}]$ and $\rho'={\rm Tr}_B[\proj{\psi'}_{AB}]$ resulting from tracing over Bob's modes of the corresponding pure states satisfy the majorization relation $\rho'\succ\rho$ (see Appendix). Using the normal form of states $\ket{\psi}_{AB}$ and $\ket{\psi'}_{AB}$, this majorization condition can be rewritten as 
\begin{equation}
\Sigma' \equiv \bigotimes_{i=1}^N\sigma_{\nu_i'} \succ   \Sigma \equiv \bigotimes_{i=1}^N\sigma_{\nu_i}   ,
\label{major-tensor}
\end{equation}
where $\sigma_{\nu_i}$ stands for a thermal state of mean photon number $\nu_i$ on the $i$-th mode. Note that the mean number of photons $\nu_i$
of state $\sigma_{\nu_i}$ is connected to the squeezing parameter $r_i$ of its parent pure state $|\Phi_{r_i}\rangle$ through the relation $\nu_i=\sinh^2(r_i)$.
Equation (\ref{major-tensor}) means that states $\Sigma$ and $\Sigma'$ admit 
respective vectors of eigenvalues $\boldsymbol{\lambda}$ and $\boldsymbol{\lambda'}$ that satisfy
the majorization relation $\boldsymbol{\lambda'}\succ\boldsymbol{\lambda}$, which is strictly equivalent to the existence of an infinite column-stochastic matrix $\mathbf{D}$ satisfying the relation $\boldsymbol{\lambda}=\mathbf{D}\boldsymbol{\lambda'}$, hence mapping state $\Sigma'$ onto state $\Sigma$.

The technical novelty of our work consists in finding a systematic way of constructing such matrices $\mathbf{D}$ for different ensembles of thermal states $\Sigma$ and $\Sigma'$. Once such a matrix $\mathbf{D}$ is found, we know that it must be possible to transform $\Sigma'$ into $\Sigma$ by using a random mixture of unitaries, i.e.,
\begin{equation}
\Sigma'=\sum_s p_s \, U_s  \Sigma U_s^\dagger,
\label{mixUni}
\end{equation}
where the unitary $U_s$ acts upon $N$ modes (on Alice's side) and is applied with probability $p_s$. (The sum over $s$ could also be replaced by an integral over 
a continuous variable and $p_s$ would then be a probability density.)
At this point, it is tempting to use standard Gaussian transformations that are well known to map thermal states onto thermal states. For instance, knowing that the mean number of photons of a thermal state is increased by applying a quantum-limited amplifier $\mathcal{A}$ (or decreased if we apply a pure-loss channel $\mathcal{L}$), one would be tempted to use $\mathcal{A}$ in order to transform a thermal state of mean photon vector $\nu'$ into output thermal states of mean photon vector $\nu$ with $\nu\ge \nu'$. Unfortunately, this Gaussian channel $\mathcal{A}$ (as well as $\mathcal{L}$) can not be written as a mixture of unitaries, so it cannot directly be used for our purposes.

However, the key observation is that the action of some specific tensor products of quantum-limited amplifiers $\mathcal{A}$ and pure-loss channels $\mathcal{L}$ on the eigenvectors of some specific tensor products of thermal states is equivalent to the action of a column-stochastic matrix $\mathbf{D}$, i.e., $\boldsymbol{\lambda}=\mathbf{D}\boldsymbol{\lambda'}$. 
This means that there must exist a mixture of (maybe non-Gaussian) unitaries having the same effect 
as the chosen tensor product of Gaussian channels when transforming the tensor product of thermal states $\Sigma'$ into the tensor product of thermal states $\Sigma$.
This implies the existence of a (maybe non-Gaussian) LOCC transformation that connects the Gaussian pure states $\ket{\psi}_{AB}\rightarrow\ket{\psi'}_{AB}$.

Below, in subsection A, we describe our approach in the simplest case of $2\times 2$ modes in the scenario \ref{a}, where the mean photon number of the two thermal states decreases (or equivalently both squeezing parameters $r_1$ and $r_2$ decrease), recovering the existence condition for a GLOCC as in \cite{Giedke}.
In subsection B, we generalize the method to give a sufficient condition for the existence of a LOCC transformation in cases \ref{c} and \ref{d}, where no GLOCC transformations are known to exist, which leads to the proof of Theorem \ref{MainResult}. In subsection C, the method is generalized to an arbitrary number of modes.






\subsection{Quantum-limited amplifier}

\begin{figure}[t]
\centering
\subfigure[]{\label{3a}\includegraphics[width=38mm]{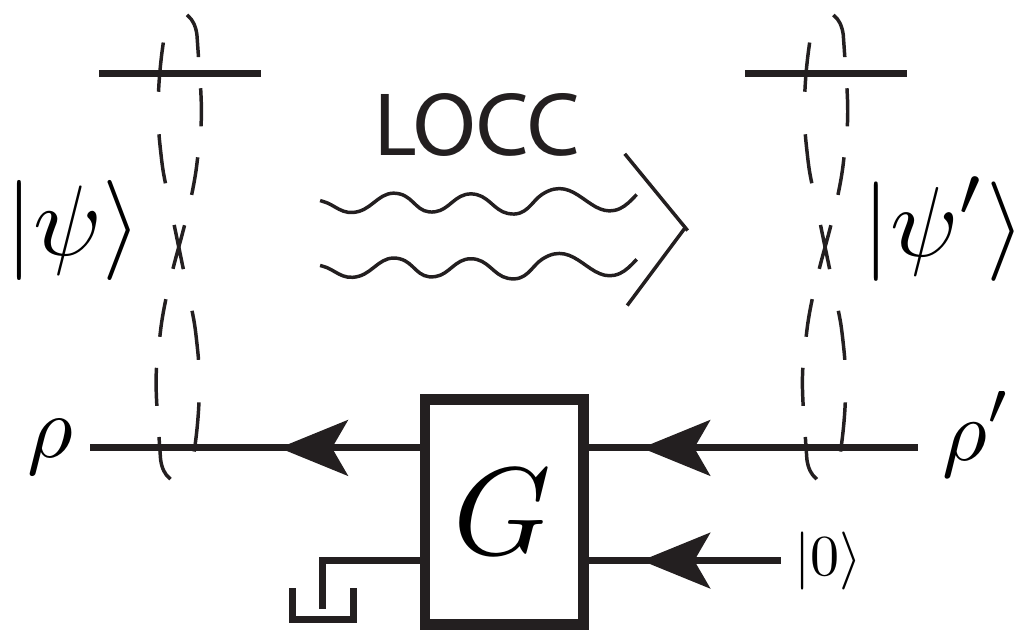}}
\hspace{1em}
\subfigure[]{\label{3b}\includegraphics[width=38mm]{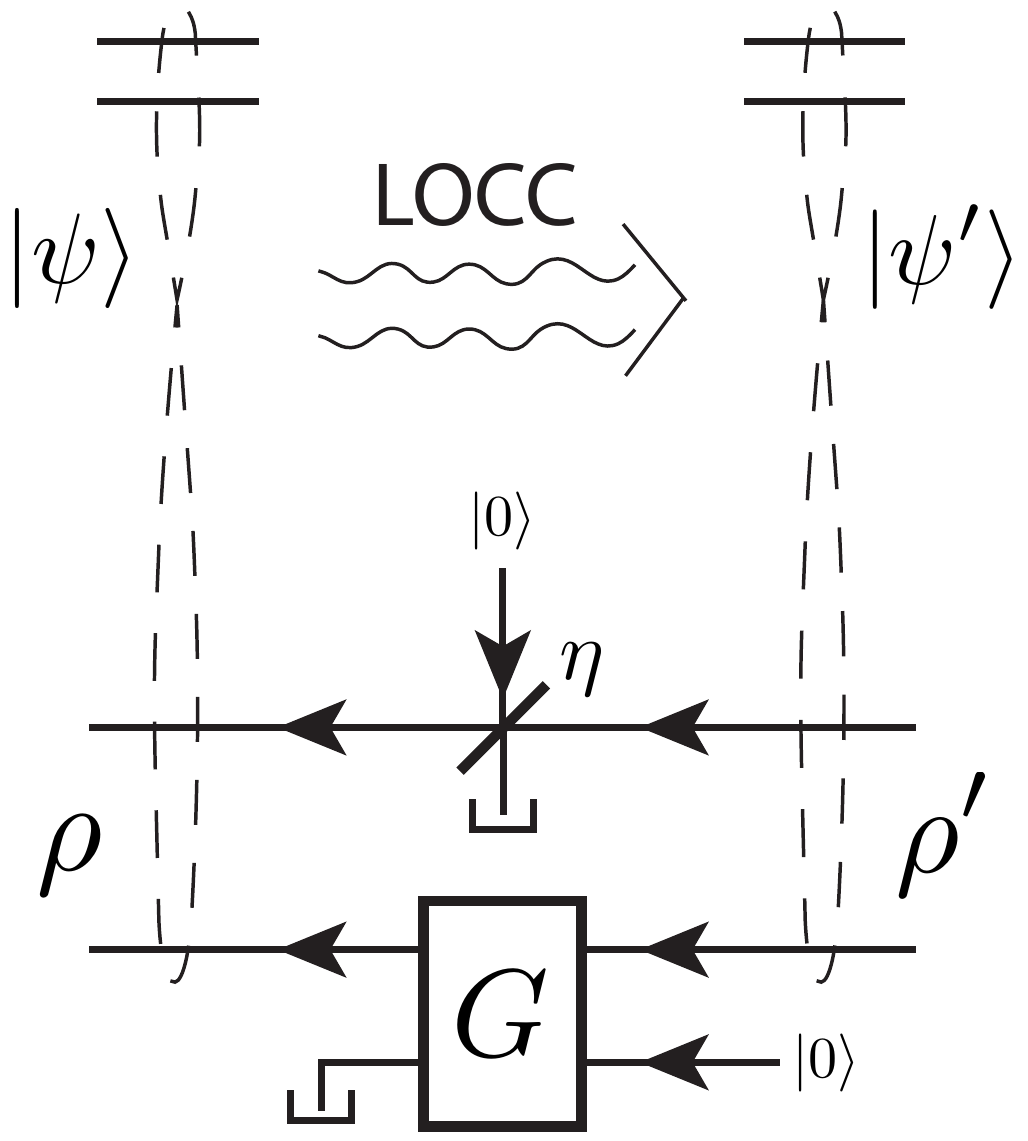}}
\caption{A bipartite pure state $\ket{\psi}$ can be transformed into $\ket{\psi'}$ with a LOCC if and only if the reduced state $\rho'$ majorizes $\rho$, or equivalently if $\rho'$ can be mapped onto $\rho$ with a column-stochastic matrix $\boldsymbol{D}$. The case of $1\times 1$ modes is shown in \ref{3a}, where matrix $\boldsymbol{D}^{\mathcal{A}}$ acts on Fock-diagonal states similarly as a quantum-limited amplifier $\mathcal{A}$ of gain $G$ ($\boldsymbol{D}^{\mathcal{A}}$ is column-stochastic provided that $G\ge 1$).  The case of $2\times 2$ modes is shown in \ref{3b}, where matrix $\boldsymbol{D}^{\mathcal{L}}\otimes\boldsymbol{D}^{\mathcal{A}}$ acts on Fock-diagonal states similarly as the product of a pure loss channel $\mathcal{L}$ of transmittance $\eta$ and a quantum-limited amplifier $\mathcal{A}$ of gain $G$ ($\boldsymbol{D}^{\mathcal{L}}\otimes\boldsymbol{D}^{\mathcal{A}}$ is column-stochastic provided that $\eta G \ge 1$). Note that  $\mathcal{L} \otimes \mathcal{A}$ is a Gaussian map, while matrix $\boldsymbol{D}^{\mathcal{L}}\otimes\boldsymbol{D}^{\mathcal{A}}$ effects a non-Gaussian transformation since condition~(\ref{eqGiedke}) is not satisfied.}
\label{fig3}
\end{figure}

As pictured in Fig.~\ref{3a}, the action of a quantum-limited amplifier $\mathcal{A}$ on an input state $\rho'$ is equivalent to
the unitary interaction between the input mode $A$ and an environmental mode $E$ initially prepared in the vacuum state $\ket{0}$
\begin{equation}
\mathcal{A} \left( \rho' \right) = \text{Tr}_E \Big[ U_{\mathcal{A}} \left( \rho' \otimes \ket{0}_E\bra{0} \right) U_{\mathcal{A}}^{\dagger} \Big],
\end{equation}
where $U_{\mathcal{A}} = \exp \left[ \frac{s}{2} \left( a_A a_E - a_A^{\dagger}a_E^{\dagger} \right) \right]$ is a two-mode squeezing unitary. When we apply the quantum-limited amplifier to a single-mode phase-invariant state $\rho' = \sum_n \lambda_n'  \ket{n}\bra{n}$ (where $\ket{n}$ is a Fock state), the output state $\rho$
is also a phase-invariant state, therefore it is also diagonal in the Fock basis, i.e., $\rho = \sum_n \lambda_n \ket{n}\bra{n}$.
The corresponding vectors of eigenvalues $\boldsymbol{\lambda'}$ and $\boldsymbol{\lambda}$ are related through the equation
$\boldsymbol{\lambda} =\boldsymbol{D}^{\mathcal{A}} \boldsymbol{\lambda'}$,
where the matrix $\boldsymbol{D}^{\mathcal{A}}$ reads
\begin{equation}
\boldsymbol{D}^{\mathcal{A}} =
\begin{pmatrix}
(1-\gamma^2) & 0 & 0 & \cdots \\
(1-\gamma^2) \gamma^2 & (1-\gamma^2)^2 & 0 & \cdots \\
(1-\gamma^2) \gamma^4 & 2 (1-\gamma^2)^2 \gamma^2 & (1-\gamma^2)^3 & \cdots  \\
\vdots & \vdots & \vdots & \ddots
\end{pmatrix}. \nonumber
\end{equation}
or in a compact form $D^{\mathcal{A}}_{n,m} = Q^{(m-1)}_{n-m} H(n-m)$ with
\begin{equation}
Q^{(i)}_{j-i} = {j \choose i} (1-\gamma^2)^{i+1} (\gamma^2)^{j-i}, \quad \gamma = \tanh(s). \nonumber
\end{equation}
and $H(x)$ being the Heaviside step function defined as $H(x)=1$ for $x \geq 0$ and $H(x)=0$ for $x < 0$.
Here, $m$ is the column index ($m-1$ is the number of input photons), $n$ is the row index ($n-1$ is the number of output photons), so that $n-m$ is the number of photons created by parametric amplification.

An important feature of this transformation $\mathcal{A}$ is that it gives a column-stochastic matrix $\boldsymbol{D}^{\mathcal{A}}$ in Fock basis. To prove this, notice that the sum of the elements of the $m$-th column of $\boldsymbol{D}^{\mathcal{A}}$ is given by
\begin{equation}
C^{\mathcal{A}}_i = \sum_{n=1}^{\infty} D^{\mathcal{A}}_{n,m} = \sum_{j=i}^{\infty} Q^{(i)}_{j-i}, \quad i=m-1.
\end{equation}
Using the Pascal identity
\begin{equation}
{n \choose k} = {n-1 \choose k-1} + {n-1 \choose k},
\label{recur}
\end{equation}
one can show that $C^{\mathcal{A}}_{i+1} = C^{\mathcal{A}}_i$ for all $i \geq 0$, which means that $C^{\mathcal{A}}_i = C^{\mathcal{A}}_0 = 1$ for all $i \geq 0$. The sum of the elements of the $n$-th row of $\boldsymbol{D}^{\mathcal{A}}$ is given by
\begin{equation}
R^{\mathcal{A}}_j = \sum_{m=1}^{\infty} D^{\mathcal{A}}_{n,m} = \sum_{i=0}^j Q^{(i)}_{j-i}, \quad j=n-1.
\end{equation}
Using the formula for the binomial series, it is trivial to see that $R^{\mathcal{A}}_j = 1 - \gamma^2 = 1/G$ for all $j \geq 0$, where $G = \cosh^2(s)$ is the intensity gain of the amplifier. Since $G \geq 1$,  it is clear that $R^{\mathcal{A}}_j \le 1$ for all $j \geq 0$, so we conclude that $\boldsymbol{D}^{\mathcal{A}}$ is indeed column-stochastic. 

This implies that there exists a set of random unitaries that maps $\rho'$ to $\rho$ and behaves exactly the same way as the quantum-limited amplifier $\mathcal{A}$ on Fock-diagonal input states (in particular, on a thermal state $\sigma$). As a consequence, the bipartite $1\times 1$ pure state $\ket{\psi}$ can be transformed into $\ket{\psi'}$ by using a deterministic LOCC, see Fig.~\ref{3a}. 
More generally, since the tensor product of column-stochastic matrices is itself column-stochastic, if each mode of $\Sigma$ is the output of a quantum-limited amplifier applied to the corresponding mode of $\Sigma'$ initially prepared in a thermal state, we have found a column-stochastic matrix which maps the vector of eigenvalues of $\Sigma'$ to that of $\Sigma$. This implies that there exists a set of random unitaries that maps $\Sigma'$ to $\Sigma$, hence the bipartite $N\times N$ pure state $\ket{\psi}$ can be transformed into $\ket{\psi'}$ by using a deterministic LOCC.

This reasoning provides a sufficient condition for the existence of a deterministic LOCC protocol in case \ref{a} of Figure \ref{fig1}, where both squeezing parameters increase. As a corollary, it also implies the non-existence of a LOCC in the reverse case \ref{b}. Thus,  our approach gives an alternative proof to Ref.~\cite{Giedke} for the existence (non-existence) of a LOCC in case \ref{a} (\ref{b}). The LOCC achieving such a transformation in case \ref{a} is rather simple: Alice combines each mode with vacuum into a beamsplitter of transmissivity $\nu_i/\nu_i'$ followed by an heterodyne detection on the environmental output port, then Alice and Bob apply displacement operations conditioned on
the outcome of the heterodyne measurement (see Supplemental material of \cite{Raul}). Hence, the entanglement transformation is achieved by a Gaussian LOCC, in accordance with condition (\ref{eqGiedke}).

\subsection{Combination of quantum-limited amplifier and pure-loss channel}

In order to study cases \ref{c} and \ref{d}, which are not covered by condition \eqref{eqGiedke}, we will now prove the existence of LOCC protocols by building column-stochastic matrices that act on the vector of eigenvalues of $\Sigma'$ similarly as the tensor product of a quantum-limited amplifier $\mathcal{A}$ and a pure-loss channel $\mathcal{L}$  would act on $\Sigma'$. In contrast with case \ref{a}, we will see here that the LOCC may be non Gaussian, even though it is based on the existence of Gaussian underlying maps $\mathcal{A}$ and $\mathcal{L}$.

As shown in Fig.~\ref{3b}, the pure-loss channel $\mathcal{L}$ acts on an input state $\rho'$ equivalently as a unitary operation acting on the input mode $A$ and an environmental mode $E$ prepared in the vacuum state $\ket{0}$, i.e.
\begin{equation}
\mathcal{E} \left( \rho' \right) = \text{Tr}_E \Big[ U_{\mathcal{E}} \left( \rho' \otimes \ket{0}_E\bra{0} \right) U_{\mathcal{E}}^{\dagger} \Big],
\end{equation}
where $U_{\mathcal{E}} = \exp \left[ \theta \left( a_A^{\dagger} a_E - a_A a_E^{\dagger} \right) \right]$ is a beam-splitter unitary. When we apply the pure-loss channel $\mathcal{L}$ to a single-mode phase-invariant state $\rho' = \sum_n \lambda_n' \ket{n}\bra{n}$, the vector of eigenvalues $\boldsymbol{\lambda}$ of the resulting state $\rho= \sum_n \lambda_n \ket{n}\bra{n}$ is related to the vector of eigenvalues $\boldsymbol{\lambda'}$ of $\rho'$ through the relation $\boldsymbol{\lambda} = \boldsymbol{D}^{\mathcal{L}} \boldsymbol{\lambda'}$ where the matrix $\boldsymbol{D}^{\mathcal{L}}$ reads
\begin{equation}
\boldsymbol{D}^{\mathcal{L}} =
\begin{pmatrix}
1 & (1-\eta) & (1-\eta)^2 & \cdots \\
0 & \eta & 2 \eta (1-\eta) & \cdots \\
0 & 0  & \eta^2 & \cdots  \\
\vdots & \vdots & \vdots & \ddots
\end{pmatrix}. \nonumber
\end{equation}
or in a more compact form $D^{\mathcal{L}}_{n,m} = P^{(m-1)}_{n-1} H(m-n)$ with
\begin{equation}
P^{(j)}_i = {j \choose i} \eta^i (1-\eta)^{j-i}, \quad \eta = \cos^2\theta. \nonumber
\end{equation}
Here again, $m$ is the column index ($m-1$ is the number of input photons), $n$ is the row index ($n-1$ is the number of output photons), so that $m-n$ is the number of photons lost in the environment.

It turns out that the matrix corresponding to this transformation $\mathcal{L}$ is column-stochastic in the trivial case $\eta=1$ only.
The sum of the elements of the $m$-th column of $\boldsymbol{D}^{\mathcal{L}}$ is given by
\begin{equation}
C^{\mathcal{L}}_j = \sum_{n=1}^{\infty} D^{\mathcal{L}}_{n,m} = \sum_{i=0}^{j} P^{(j)}_i, \quad j=m-1.
\end{equation}
Using the binomial series formula, it is straightforward to see that $C^{\mathcal{L}}_j = 1$ for all $j \geq 0$. This is consistent with the fact that the elements of $\boldsymbol{\lambda}$ should form a probability distribution, regardless of  $\boldsymbol{\lambda'}$ (the map $\mathcal{L}$ conserves the normalization of probabilities). However, the sum of the elements of the $n$-th row of $\boldsymbol{D}^{\mathcal{L}}$ is given by
\begin{equation}
R^{\mathcal{L}}_i = \sum_{m=1}^{\infty} D^{\mathcal{L}}_{n,m} = \sum_{j=i}^{\infty} P^{(j)}_i, \quad i=n-1.
\end{equation}
Using again identity \eqref{recur}, one can prove that $R^{\mathcal{L}}_{i+1} = R^{\mathcal{L}}_i$ for all $i \geq 0$, meaning that $R^{\mathcal{L}}_i = R^{\mathcal{L}}_0 = 1/\eta$ for all $i \geq 0$. Since the intensity transmittance $\eta$ of the pure-loss channel satisfies $0 \leq \eta \leq 1$, $\boldsymbol{D}^{\mathcal{L}}$ is column-stochastic only if $\eta = 1$.

Despite $\boldsymbol{D}^{\mathcal{L}}$ not being column-stochastic, the tensor product of a quantum-limited amplifier and a pure-loss channel may still provide us with useful column-stochastic transformations, as indicated in the following theorem [see also Fig.~\ref{3b}]:

\begin{theo}
Let $\rho$ and $\rho'$ be tensor products of two thermal states. If it is possible to transform $\rho'$ into $\rho$ by applying a tensor product of a pure-loss channel $\mathcal{L}$ characterized by $\eta$ and a quantum-limited amplifier $\mathcal{A}$ characterized by $G$ such that $\eta G \geq 1$, then $\rho' \succ \rho$.
\label{Main2modes}
\end{theo}

The proof uses the fact that the sum of the row elements (column elements) of a tensor product of two matrices
is given by the product of a proper selection of sums of row elements (colum elements) of the component matrices.
Since the columns of $\boldsymbol{D}^{\mathcal{L}}$ and $\boldsymbol{D}^{\mathcal{A}}$ sum to $1$, while their rows sum to 
$1/\eta$ and $1/G$, respectively, it is easy to see that the matrix $\boldsymbol{D}^{\mathcal{L}}\otimes\boldsymbol{D}^{\mathcal{A}}$ is such that 
its columns sum to $1$ and its rows sum to $1/(\eta G)$. Therefore, the matrix $\boldsymbol{D}^{\mathcal{L}}\otimes\boldsymbol{D}^{\mathcal{A}}$ 
is column-stochastic if $\eta G \geq 1$ $\Box$.


Now, Theorem \ref{MainResult} can be easily proven by using Theorem \ref{Main2modes}. As explained in the Appendix, the LOCC transformation $\ket{\psi} \rightarrow \ket{\psi'}$ is possible iff the majorization relation $\rho' \succ \rho$ holds. According to Theorem \ref{Main2modes}, this is possible if we can apply a pure-loss channel $\mathcal{L}$ to one of the modes of $\rho'$ and a quantum-limited amplifier $\mathcal{A}$ to the other mode of $\rho'$ such that condition $\eta G \geq 1$ is verified. What is surprising here is that  $\mathcal{L}\otimes \mathcal{A}$ is a Gaussian map, but we use it to ensure the existence of a non-Gaussian transformation from $\rho'$ to $\rho$, guaranteeing in turn the existence of a non-Gaussian LOCC transformation from $\ket{\psi}$ to $\ket{\psi'}$.

Let us assign a subscript $\eta$ to the mode whose squeezing parameter increases ($r_\eta'  >  r_\eta$) along the transformation  $\ket{\psi} \rightarrow \ket{\psi'}$, and a subscript $G$ to the mode whose squeezing parameter decreases ($r_G'  <  r_G$). 
When a pure-loss channel is applied on a thermal state of mean photon number $\nu'_{\eta}$, it is transformed into another thermal state of mean photon number $\nu_{\eta} = \eta \nu'_{\eta}$. Similarly, when a quantum-limited amplifier is applied on a thermal state of mean photon number $\nu'_G$, it is transformed into another thermal state of mean photons number $\nu_G = G \nu'_G + (G-1)$. Condition $\eta G \geq 1$ therefore becomes
\begin{equation}
\frac{\nu_{\eta} (\nu_G+1)}{\nu'_{\eta} (\nu'_G+1)} \geq 1 \Leftrightarrow \frac{\sinh(r_{\eta}) \cosh(r_G)}{\sinh(r'_{\eta}) \cosh(r'_G)} \geq 1.
\end{equation}
Using the properties of the hyperbolic functions, one can easily prove that this is equivalent to
\begin{equation}
\frac{\sinh(r_{\eta}+r_G) + \sinh(r_{\eta}-r_G)}{\sinh(r'_{\eta}+r'_G) + \sinh(r'_{\eta}-r'_G)} \geq 1.
\end{equation}
What differentiates cases \ref{c} and \ref{d} is whether it is the squeezing parameter of the first or second mode that increases along transformation  $\ket{\psi} \rightarrow \ket{\psi'}$.
In case \ref{c}, $r_1$ increases ($r_2$ decreases), while the reverse holds in case \ref{d}. Thus, the roles of the quantum-limited amplifier and pure-loss channel are exchanged between these two cases. In case \ref{c}, $r_{\eta}=r_1$ and  $r_G=r_2$, so we recover Eq.~(\ref{eqMain}) with the plus sign, consistent with $r_1'-r_1\ge  0$. In case \ref{d}, 
$r_{\eta}=r_2$ and  $r_G=r_1$, so we recover Eq.~(\ref{eqMain}) with the minus sign, consistent with $r_1'-r_1< 0$. 
This concludes the proof of Theorem \ref{MainResult}.



\subsection{Extension to an arbitrary number of modes}

It is easy to see, from the derivation of Theorem \ref{Main2modes}, that it can be extended to the $N\times N$ case, with $N$ being an arbitrary number of modes. We obtain the following generalization:

\begin{theo}
Let $\rho$ and $\rho'$ be tensor products of $N$ thermal states characterized by vectors of mean number of photons $(\nu_1,\cdots \nu_N)$ and $(\nu'_1,\cdots \nu'_N)$, respectively. If it is possible to transform $\rho'$ into $\rho$ by applying a tensor product of quantum-limited Gaussian channels $\mathcal{C}_k$, where
channel $\mathcal{C}_k$ acting on mode $k$ is either a pure-loss channel or a quantum-limited amplifier, and where the set of channels $\mathcal{C}_k$ is such that
\begin{equation}
\prod_{k=1}^N \tau_k \geq 1
\label{prodTauGeq1}
\end{equation}
\begin{equation}
\text{with} \quad \tau_k = \left\lbrace \begin{aligned}
& \eta_k \quad \text{if $\mathcal{C}_k$ is a pure-loss channel} \\
&G_k \quad \text{if $\mathcal{C}_k$ is a quantum-limited amplifier}
\end{aligned} \right. \nonumber
\end{equation}
 then $\rho' \succ \rho$.
\label{MainNmodes}
\end{theo}

The transformation described in Theorem \ref{MainNmodes} reads
\begin{equation}
\begin{aligned}
\rho' = & \sigma_{\nu_1'} \otimes \sigma_{\nu_2'} \otimes \cdots \otimes \sigma_{\nu_N'} \\
& \left\downarrow\rule{0cm}{.4cm}\right. \mathcal{C}_1 \ \ \left\downarrow\rule{0cm}{.4cm}\right. \mathcal{C}_2 \ \ \ \cdots \ \ \left\downarrow\rule{0cm}{.4cm}\right. \mathcal{C}_n \\
\rho = & \sigma_{\nu_1} \otimes \sigma_{\nu_2} \otimes \cdots \otimes \sigma_{\nu_N} 
\end{aligned} \nonumber
\label{multimode}
\end{equation}
and the proof goes the same way as in the $2\times 2$ case, exploiting the fact that the sum of row elements (column elements)
of a tensor product of matrices is given by the product of a proper selection of sums of row elements (columns elements) of the individual matrices.

If condition (\ref{prodTauGeq1}) is satisfied, then $\rho'$ majorizes $\rho$, which in turn implies that any purification of $\rho$ (noted $\ket{\psi}$) can be mapped into a purification of $\rho'$ (noted $\ket{\psi'}$) using a deterministic LOCC, for any number of modes $N$.
As in the $2\times 2$ case, this theorem result generalizes the sufficient condition in Eq.~\eqref{eqGiedke} when
all squeezing parameters $(r_1,\cdots r_N)$ increase, in which case a GLOCC transformation works. 
Again, for more complicated evolution patterns of the squeezing parameters $(r_1,\cdots r_N)$ where Eq.~\eqref{eqGiedke} precludes the existence of a GLOCC,
our Theorem \ref{MainNmodes} may very well permit a non-Gaussian LOCC to achieve the transformation. We have not systematically explored all possibilities for the evolution pattern 
$(r_1,\cdots r_N)\rightarrow (r_1',\cdots r_N')$, but it is clear that Theorem \ref{MainNmodes} gives a sufficient condition for LOCC transformations $\ket{\psi} \rightarrow \ket{\psi'}$ that encompasses situations that are not covered by condition~\eqref{eqGiedke}.



\section{Conclusion}

In this work, we have developed a technique to find existence conditions on entanglement transformations between bipartite pure Gaussian states that go beyond Gaussian local operations and classical communication (GLOCC). We have presented a new sufficient criterion for the existence of a deterministic LOCC transforming a pure $N\times N$ Gaussian state into another. This result generalizes Giedke \textit{et al.}'s necessary and sufficient criterion for the existence of a GLOCC relating such pure Gaussian states  \cite{Giedke} (while loosing the necessary character of the condition, meaning that a LOCC transformation may exist that is not detected by our criterion). In particular, our criterion guaranties the existence of a non-Gaussian LOCC connecting some pure Gaussian states that cannot be connected otherwise with a GLOCC according to Ref.  \cite{Giedke}. In other words, we exhibit situations where pure Gaussian state interconversions can be achieved with non-Gaussian local operations even though Gaussian local operations alone cannot. This is reminiscent of situations where a Gaussian no-go theorem precludes the use of Gaussian resources in order to achieve a task involving Gaussian states, e.g. quantum entanglement distillation \cite{Eisert-nogo,Giedke-nogo,Fiurasek}, quantum error correction \cite{Niset}, or quantum bit commitment \cite{Magnin}.

Our approach relies on majorization theory (extended to infinite-dimensional spaces) and consists in building explicit column-stochastic matrices $\boldsymbol{D}$ that map the state $\rho'$ (whose purification is $\ket{\psi'}$) onto the state $\rho$ (whose purification is $\ket{\psi}$), hence ensuring that the transformation $\ket{\psi} \rightarrow \ket{\psi'}$ is possible under a LOCC (even if a GLOCC may not suffice). We build our column-stochastic matrices $\boldsymbol{D}$ by using (tensor products of) Gaussian channels (namely, $\mathcal{A}$ and $\mathcal{L}$) applied to Fock-diagonal states. Thus, ironically, our approach allows us to infer the existence of non-Gaussian LOCC transformations without leaving the simple mathematical tools developed for Gaussian channels. 
Unfortunately, working with an infinite-dimensional space makes it highly non-trivial to find the actual set of random unitaries  mapping $\rho'$ to $\rho$ and thus to design the corresponding LOCC protocol, even while $\mathbf{D}$ is known. Hence, the tools developed for finite-dimensional spaces cannot be easily applied in the quantum optical scenario considered here.
We leave this subject for further investigation.

To simplify our presentation, we have restricted our numerical analysis to the $2 \times 2$ case (see Fig.~\ref{fig2}). A thorough analysis of 
the entanglement transformations between $N\times N$ states would be worthwhile, but we already observe interesting behaviors in the $2 \times 2$ case (situations where GLOCCs are precluded by  \cite{Giedke} while LOCCs are sufficient according to Theorem \ref{MainResult}). 
Even in the $2 \times 2$ case, a problem left open here is that of designing the specific non-Gaussian LOCC realizing the transformation that is predicted to exist.
Interestingly, the situation considered in case \ref{c} of Figure \ref{fig1} 
can be seen as a way of concentrating the entanglement of two two-mode squeezed vacuum states into a single two-mode squeezed vacuum state (the other one loosing its entanglement). Therefore, progress on designing such LOCC protocols could open a way to novel protocols enhancing the entanglement of Gaussian states through non-Gaussian operations.

\acknowledgments

This work was supported by the F.R.S.-FNRS under the ERA-Net CHIST-ERA project HIPERCOM and by the Belgian Federal IAP program under Project No. P7/35 Photonics@be. 
R.G.-P. acknowledges financial support from a Back-to-Belgium grant from the Belgian Federal Science Policy.

\appendix*

\section{Theory of majorization}

Majorization is an algebraic theory which provides a mean to compare two probability distributions in terms of disorder or randomness \cite{Maj1903, Ineq}. Let $\mathbf{p}$ and $\mathbf{q}$ be two probability distribution vectors of dimension $n$. If $\mathbf{p^{\downarrow}}$ and $\mathbf{q^{\downarrow}}$ are vectors containing the elements of $\mathbf{p}$ and $\mathbf{q}$ sorted in non-increasing order, $\mathbf{p}$ majorizes $\mathbf{q}$, i.e $\mathbf{p} \succ \mathbf{q}$, iff
\begin{equation}
\sum_{i=k}^k p^{\downarrow}_i \geq \sum_{i=k}^k q^{\downarrow}_i, \quad k=1,...,n,
\label{majo1}
\end{equation}
with equality when $k=n$. In this case, one says that $\mathbf{q}$ is more disordered than $\mathbf{p}$. Note that if $\mathbf{p}$ and $\mathbf{q}$ are probability distributions, the equality for $k=n$ is always satisfied. Majorization only provides a pre-order, in the sense that if $\mathbf{p} \nsucc \mathbf{q}$, this doesn't necessarily mean that $\mathbf{p} \prec \mathbf{q}$. When both $\mathbf{p} \nsucc \mathbf{q}$ and $\mathbf{p} \nprec \mathbf{q}$ are satisfied, $\mathbf{p}$ and $\mathbf{q}$ are said to be incomparable.

In order to understand why majorization allows one to compare probability distributions in terms of disorder, let us  introduce an alternative way of detecting majorization. A more intuitive definition is to say that $\mathbf{p}$ majorizes $\mathbf{q}$ iff there exists a set of $n$-dimensional permutation matrices $\mathbf{\Pi}_n$ and a probability distribution $\left\lbrace t_n \right\rbrace$ such that
\begin{equation}
\mathbf{q} = \sum_n t_n \mathbf{\Pi}_n \mathbf{p}.
\end{equation}
This last equation clearly show the relation between disorder and majorization. Indeed, we see that if $\mathbf{p}$ majorizes $\mathbf{q}$, then $\mathbf{q}$ can be obtained by applying random permutations to $\mathbf{p}$, making the latter more disordered. This definition also allows us to introduce another equivalent way of characterizing majorization in terms of doubly-stochastic matrices. These are the matrices whose columns and rows sum to $1$. The set of doubly-stochastic matrices of a given dimension is convex, and its extremal points are given by permutation matrices of the same dimension. Consequently, any doubly stochastic matrix can be decomposed as a convex combination of permutation matrices. This allows us to introduce the following theorem.

\begin{theo}
Given the vectors $\mathbf{p},\mathbf{q} \in \mathbb{R}^d$, $\mathbf{p} \succ \mathbf{q}$ iff
\begin{equation}
\mathbf{q} = \mathbf{D} \mathbf{p}
\end{equation}
for some doubly stochastic matrix $\mathbf{D}$.
\label{majo2}
\end{theo}

This theory of majorization beautifully extends to the quantum realm. Indeed, one can compare density matrices by simply comparing their vectors of eigenvalues, whose elements are probability distributions. Thus, given two density matrices $\rho$ and $\sigma$ whose vectors of eigenvalues are respectively given by vectors $\boldsymbol{\lambda}(\rho)$ and $\boldsymbol{\lambda}(\sigma)$, one says that $\rho \succ \sigma$ if $\boldsymbol{\lambda}(\rho) \succ \boldsymbol{\lambda}(\sigma)$. We then naturally have the following theorem \cite{NielsenVidal}.

\begin{theo}
$\rho \succ \sigma$ iff state $\sigma$ can be obtained from state $\rho$ by applying a random mixture of unitaries, i.e
\begin{equation}
\sigma = \sum_i t_i U_i \rho U^{\dagger}_i,
\end{equation}
where $\left\lbrace t_i \right\rbrace$ is a probability distribution and the $U_i$ are unitaries for all $i$.
\end{theo}

A very interesting connection between quantum information theory and majorization resides in the fact that one can use the latter in order to compare pure bipartite entangled state, or more precisely to investigate the possibility to transform a state into another using a LOCC. Suppose Alice and Bob share a pure state $\ket{\psi}$ and want to transform it into a state $\ket{\phi}$. The following theorem investigates such a possibility \cite{Nielsen, NielsenVidal}:

\begin{theo}
State $\ket{\psi}$ can be converted deterministically into state $\ket{\phi}$ using LOCC iff $\rho_{\psi} \prec \rho_{\phi}$, where $\rho_{\psi}$ is the reduced density matrix of system A $\rho_{\psi} \equiv \text{Tr}_B \left(\ket{\psi} \bra{\psi} \right)$ and similarly for $\rho_{\phi}$.
\label{LOCC}
\end{theo}

The theory of majorization nicely adapts to the infinite dimensional case, allowing one to compare Gaussian states in particular. The definitions we stated before stay the same, the only difference residing in the doubly stochastic matrix, which should now be replaced by an infinite dimensional column stochastic matrix, whose columns still sum to $1$, but whose rows sum to a value less or equal than $1$ \cite{infMaj}. Note that in the case of Gaussian states, it is difficult to use definition \eqref{majo1}, due to the complexity of the problem of ordering the eigenvalues of a multi-mode Gaussian state. Verifying that Theorem \ref{majo2} holds seems an easier task, which is the technique used in our work. 
Unfortunately, there is not easy algorithm to decide whether a column-stochastic matrix exists that connects the eigenvalues of two infinite sets (and that generates the matrix in case it exists).
Therefore, heuristic approaches, such as the one developed in our work, are needed. In this paper, we provide a way to find such a 
family of column-stochastic matrices, allowing us to use the theory of majorization to compare Gaussian multi-mode states.

\end{document}